# The regulation of online political micro-targeting in Europe

**Tom Dobber**
*Department of Political Communication & Journalism, University of Amsterdam, Netherlands, T.Dobber@uva.nl*

**Ronan Ó Fathaigh**
*Institute for Information Law, University of Amsterdam, Netherlands, R.F.Fahy@uva.nl*

**Frederik J. Zuiderveen Borgesius**
*Digital Security group (DiS), Radboud University, Nijmegen, Netherlands, frederikzb@cs.ru.nl*

**Abstract:** In this paper, we examine how online political micro-targeting is regulated in Europe. While there are no specific rules on such micro-targeting, there are general rules that apply. We focus on three fields of law: data protection law, freedom of expression, and sector-specific rules for political advertising; for the latter we examine four countries. We argue that the rules in the General Data Protection Regulation (GDPR) are necessary, but not sufficient. We show that political advertising, including online political micro-targeting, is protected by the right to freedom of expression. That right is not absolute, however. From a European human rights perspective, it is possible for lawmakers to limit the possibilities for political advertising. Indeed, some countries ban TV advertising for political parties during elections.

*This paper is part of <u>Data-driven elections</u>, a special issue of* Internet Policy Review *guest-edited by Colin J. Bennett and David Lyon.*

# INTRODUCTION

A new form of political advertising has emerged: online political micro-targeting ('micro-targeting'). Such micro-targeting typically involves monitoring people's online behaviour, and using the collected data, sometimes enriched with other data, to display individually targeted political advertisements. However, micro-targeting poses serious risks, as demonstrated by the Cambridge Analytica scandal, where a voter-profiling company had harvested private information from the Facebook profiles of more than 50 million users without their permission (Guardian, 2019).

Unlike political advertising on television, micro-targeting not only affects the democratic process, but it also affects people's privacy and data protection rights. Indeed, micro-targeting affects myriad other rights and duties, including a political party's and online platform's right to impart information, a voter's right to receive information, and the government's duty to ensure free and fair elections.

We focus on the following, legal, question: How is micro-targeting regulated in Europe? We examine the question from three perspectives, namely data protection law, freedom of expression (the right to receive and impart information), and sector-specific rules for political advertising. We focus on the region of the European Union, and also draw upon case law of the Council of Europe's European Court of Human Rights.

First, we discuss the General Data Protection Regulation, which lays down rules on the use of personal data. Second, we examine how political parties enjoy a freedom of expression claim regarding their advertising. We discuss, among other things, whether a political party's freedom of expression gives such a party the right to choose its advertising medium.

Finally, many countries have sector-specific rules for political advertising, which differ from country to country. By way of illustration, we discuss the rules in Germany, France, the Netherlands, and the UK. For decades, paid political advertising on television has been completely banned during elections in many European democracies. These political advertising bans aim to prevent the distortion of the democratic process by financially powerful interests, and to ensure a level playing field during elections. But before we discuss regulation, we give a brief introduction to online political micro-targeting.

# MICRO-TARGETING

Online political micro-targeting, or micro-targeting for short, can be summarised as consisting of three steps: 1) collecting personal data, 2) using those data to identify groups of people that are likely susceptible to a certain message, and 3) sending tailored online messages (Zuiderveen Borgesius et al., 2018). The objective of micro-targeting can be manifold: to persuade, inform, or mobilise, or rather to dissuade, confuse or demobilise voters. People can be micro-targeted on the basis of all kinds of information (such as their personality traits, their location, or the issues they care about). Hence, any data can be valuable: from consumer data to browsing behaviour. Such data can provide enough information to make inferences about the susceptibilities of the target audiences.

*Micro-targeting* differs from regular targeting not necessarily in the size of the target audience,

but rather in the level of homogeneity, perceived by the political advertiser. Simply put, a *micro*-targeted audience receives a message tailored to one or several specific characteristic(s). This characteristic is perceived by the political advertiser as instrumental in making the audience member susceptible to that tailored message. A *regular* targeted message does not consider matters of audience heterogeneity.

For example, the Green Party plans to target a neighbourhood in Amsterdam. The party chooses this specific neighbourhood and not the adjacent neighbourhood, because the city's statistics show that turnout was low last election but the number of votes for the Green Party was high. The Green Party sends a political message to everyone living in that neighbourhood. We would classify this as *regular* targeting.

The Green Party would be *micro-targeting* when it acknowledges that the neighbourhood consists of many people that may share socio-demographics, but they still have many different reasons to vote for a specific party. Some want cheap solar panels, others want more nature in the city, others want to block cars from the city centre, others want a softer stance on immigration, drugs, etc. Moreover, some people in the neighbourhood would never vote and others would never vote for the Green Party. When *micro-targeting*, the Green Party could ignore the unlikely voters and tailor their messages to possible voters' issue salience (or other characteristics). This way the Green Party would turn one heterogeneous group into several homogeneous subgroups.

To illustrate micro-targeting in practice: in the Netherlands, almost all political parties use Facebook's *lookalike audiences* function to micro-target voters (Dobber et al., 2017). Political parties use this function to find people who fit a very specific profile, for example, party members that share a (or more) specific characteristic(s).

Dutch pro-immigrant party DENK took an innovative approach when they micro-targeted only the people who use a special sim card. This sim card can be used to cheaply call non-EU countries. In practice, mostly immigrants use those sim cards, giving DENK a simple way to efficiently reach people who are traditionally difficult to reach (Van Trigt, 2018). DENK is known to have experimented with fear appeals, meant to scare their own base to the polls (a false advertisement made to look like it came from Geert Wilders' Freedom Party, with the statement: "after March 15 [election day] we are going to cleanse the Netherlands"). Such fear appeals can be easily distributed to people who own the special sim cards (Nieuws BV, 2018).

Micro-targeting techniques develop quickly, and so do the ways in which political actors employ them (Kreiss and Barrett, 2019). In the pre-mass media age, citizens received a 'micro-targeted' message, or at least a personalised one, when the local cleric visited his parish's homes to remind them why and for which party they should vote (Kaal, 2016). The advent of the internet and social media in particular enabled micro-targeting on a much larger scale than in the pre-mass media age. Moreover, the cost in time and effort is much lower, and the variation in messages can be enormous. In addition, people often do not know they have been micro-targeted (and *if* they do, they remain in the dark about what kind of information was used, although Facebook does provide some information about the targeting criteria specified by the advertiser), while that was clear when the cleric knocked on the door. Back then, for instance, you could act as if you were not home. It is more difficult to escape micro-targeted political messages. People leave behind their data at every move they make. Consequently, they can be targeted at *any* moment, with increasing precision (Zuiderveen Borgesius et al., 2018).

Micro-targeting originates from the United States, where relatively loose data-protection regulation may have facilitated the rapid development and adoption of the technique (Bimber, 2014; Kreiss, 2012, 2016; Nickerson & Rogers, 2014; Nielsen, 2012). However, micro-targeting is gaining traction in the EU. For example, micro-targeting was used for the first time on a large scale in national elections in the UK (Anstead, 2017), the Netherlands (Dobber et al., 2017), Germany (Drepper, 2017), and France (Liegey Muller Pons, n.d.; International IDEA, 2018).

Facebook, due to its easy-to-use infrastructure makes it easy for EU parties to use micro-targeting. Facebook and Google hold vast amounts of personal data and offer political parties the means to reach specific groups without having to collect data. Naturally, micro-targeting in the EU does not solely occur on Facebook. Political parties can also develop micro-targeting techniques by themselves, or they can, for instance, hire the services of specialised firms.

While micro-targeting is gaining popularity with political parties throughout Europe, the use of the technique brings risks. Micro-targeting poses risks to individuals, political parties, and public opinion (Zuiderveen Borgesius et al., 2018). For *individuals*, micro-targeting threatens privacy. For example, a data breach could expose information about individuals' income, education, consumer behaviour, but also their inferred political leanings, sexual preferences, or religiosity. Cambridge Analytica harvested the data of tens of millions of unwitting voters (Cadwalladr & Graham-Harrison, 2018). Moreover, merely being aware of data collection could evoke chilling effects: people may alter their behaviour if they suspect being under surveillance (Richards, 2015; Dobber et al., 2018). Manipulation is a different risk to individuals (see Susser, Roessler, & Nissenbaum, 2019). The 2016 US elections saw disinformation efforts targeted, for example, to African Americans (Howard, Ganesh, & Liotsiou, 2018). Finally, political actors could ignore certain voter groups they deem unimportant ('redlining', see Howard, 2006) or demobilise the supporters of competing parties (Green & Issenberg, 2016). A consequence could be underrepresentation of certain societal groups.

The costs of micro-targeting and the power of digital intermediaries are among the main risks to *political parties*. The costs of micro-targeting may give an unfair advantage to the larger and better-funded parties over the smaller parties. This unfair advantage worsens the inequality between rich and poor political parties (see Margolis & Resnick, 2000), and restrains the free flow of political ideas. Second, digital intermediaries profit from their vast amounts of personal data and their intuitive infrastructure. Political parties are dependent on these intermediaries to run a modern political campaign.

On the level of public opinion, micro-targeting makes it difficult to find out which issues candidates find most important, and which they least care about. Moreover, an elected official may have trouble interpreting her mandate when a large range of issues was covered during a political campaign. Finally, micro-targeting could lead to a fragmentation of the marketplace of ideas. Fragmentation happens when the public loses track of overarching themes, and instead focuses on the single issues that are relevant to them personally, which are the topics delivered through micro-targeting techniques (Hillygus & Shields, 2008; Zuiderveen Borgesius et al., 2018).

Advancements in technology lead to increasing possibilities to influence voters' behaviour. Micro-targeting can be an important tool for (foreign) political actors to interfere in elections. Think of micro-targeted deep fakes (manipulated, but realistic, videos) that can be used to misinform specific voter groups. Malicious political actors can use micro-targeting to reach the right voter with the right disinformation message, thereby maximising the impact of each specific message (Bayer et al., 2019).

Many contextual factors play a role in shaping micro-targeting. For instance, the electoral system is important (Dobber et al., 2017). A political advertiser operating in a multiparty system makes different choices than an advertiser operating in a (*de facto*) two-party system. A country's, or an electoral district's culture or tradition also plays a role. When there is a low turnout culture, for instance, political advertisers focus more on getting out the vote than on persuading voters. And US campaigns frequently engage in attack ads (Vafeiadis, Li, & Shen, 2018), while attack ads are rare in, for instance, Japan (Plasser & Plasser, 2002). In addition, the campaign team level, resource factors, organisational factors, infrastructural factors, structural electoral factors (Kreiss, 2016), and ethical and legal concerns play a role in shaping micro-targeting (Dobber et al., 2017; see also Kruschinski & Haller, 2017). However, because of length constraints, this paper focuses on how the law regulates micro-targeting.

## PRIVACY AND DATA PROTECTION RULES

Micro-targeting entails the use of personal data for targeted advertising, and therefore the applicable privacy and data protection rules are relevant. The EU grants the right to the protection of personal data the status of a human right. The Charter of Fundamental Rights of the European Union (2000) includes a separate right to the protection of personal data, in addition to a general right to privacy.

Almost 25 years ago, the EU adopted the influential Data Protection Directive (1995). The EU replaced the 1995 Directive with the General Data Protection Regulation (2016; in application since 2018). The GDPR is a legal instrument that aims to ensure that personal data are only used fairly and transparently. The GDPR imposes obligations on organisations that use personal data (data controllers) and grants rights to people whose personal data are used (data subjects). Compliance with the GDPR is overseen by independent Data Protection Authorities (DPAs).

The scope of the GDPR is wide. The GDPR applies to the 'processing' of 'personal data'. Almost anything that can be done with personal data falls within the processing definition. The personal data definition also has a wide scope, and covers, for instance, tracking cookies, IP addresses, and other online identifiers (article 4(1) GDPR; Court of Justice of the European Union 2017). In many cases, the GDPR also applies to data controllers established outside the EU, for instance when they process personal data and offer goods or services to people in the EU, or when they track the online behaviour of people in the EU (article 3 GDPR).

The data protection principles that lie at the core of the GDPR (article 5), sometimes called Fair Information Principles, did not change much in comparison to the 1995 Directive. More than 120 countries in the world have data privacy laws with similar principles (Greenleaf 2017). Below we summarise, roughly, some main points of the GDPR (for more details see Hoofnagle, Van der Sloot, & Zuiderveen Borgesius, 2019).

The data protection principles that form the core of the GDPR (article 5) can be summarised as follows: (a) personal data may only be used lawfully, fairly and in a transparent manner ('lawfulness, fairness and transparency'); (b) personal data may only be collected for purposes that are specified in advance. And such data may not be used for random other purposes ('purpose limitation'); (c) controllers may not collect or use more personal data than is necessary for the processing purpose ('data minimisation'). (d) controllers must generally ensure that the personal data they use are accurate ('accuracy'); (e) personal data may not be retained for unreasonably long periods ('storage limitation'); (f) data security must be ensured ('integrity and confidentiality'); and (g) the data controller is responsible for compliance ('accountability').

Data subjects have several rights under the GDPR. For example, data subjects can demand a controller to tell them what personal data it holds on them (article 15). To illustrate: a US citizen, David Carroll, used his access rights under the Data Protection Act in the UK to obtain more information about which data the micro-targeting firm Cambridge Analytica held on him (Carroll, 2018).

The most important change brought by the GDPR is that it empowers DPAs with serious enforcement possibilities. Controllers that breach the GDPR's rules can be fined up to 20 million Euros, or up to 4% of their worldwide turnover – that is income, not profit (article 83). The mere possibility of fines has led many companies and other organisations to improve their data practices.

The GDPR does not contain specific rules for micro-targeting. The GDPR is extra strict, however, for many types of sensitive data ('special categories of personal data', article 9). Personal data regarding people's 'political opinions' fall within that category.

In principle, processing of such special categories of personal data is prohibited, but the GDPR includes exceptions to that prohibition. Political parties (and similar not-for-profit bodies) can, under certain circumstances, rely on an exception to the ban on using sensitive data. Again, the conditions are strict. For example, a political party may only use personal data of members or former members who are in regular contact with it, under certain circumstances (GDPR, article 9(2)(d)). The exception is phrased as follows.

> Paragraph 1 [the ban on using special categories of personal data] shall not apply if one of the following applies: (…) processing is carried out in the course of its legitimate activities with appropriate safeguards by a foundation, association or any other not-for-profit body with a political, philosophical, religious or trade union aim and on condition that the processing relates solely to the members or to former members of the body or to persons who have regular contact with it in connection with its purposes and that the personal data are not disclosed outside that body without the consent of the data subjects (GDPR, article 9(2)(d)).

Another possibly relevant exception is the data subject's 'explicit consent'. For targeted marketing (not conducted by political parties themselves), such explicit consent is the only available exception to the processing ban on sensitive data. The GDPR's requirements for valid consent are strict. For instance, the GDPR does not accept opt-out systems (that assume that people consent if they fail to object). And burying a consent request in the small print of a privacy notice is not allowed (article 4(11); article 7 GDPR). There are more exceptions to the ban on using sensitive data, but those exceptions are not relevant for elections.

Apart from the GDPR, the EU has separate rules for tracking cookies and similar tracking technologies (EU ePrivacy Directive, 2009). Roughly summarised, anybody who wants to set tracking cookies on somebody's computer must ask that person for his or her prior informed consent. Hence, a company that wants to use tracking cookies to trace somebody's online behaviour to learn more about that person's interests is only allowed to do so after asking consent (EU ePrivacy Directive, 2009). The EU is busy revising the rules for online tracking (Zuiderveen Borgesius et al., 2017).

The GDPR only entered into force in May 2018. From the perspective of micro-targeting technology, that is a long time ago. But for a law, the GDPR is young. Therefore, the exact meaning of many GDPR rules (including the rules on sensitive data) still has to emerge from case law.

Nevertheless, it seems likely that, when compared to the US, Europe's privacy rules hinder micro-targeting. For example, because of Europe's privacy rules, it is harder for political parties to buy data about people (see also Bennett, 2016). And in most countries in Europe, it is impossible to access voter registration records. The GDPR's transparency requirements can help journalists and researchers to find out more about what political parties and marketing companies do with personal data.

In sum, Europe's privacy laws do not categorically prohibit micro-targeting. Still, Europe's privacy laws make micro-targeting more difficult than in, for instance, the US.

The GDPR does not and will not solve all privacy problems. Compliance and enforcement leave something to be desired. And there are weak points in the GDPR, when applied to micro-targeting. For example, the GDPR is an omnibus law, applying to almost all usage of personal data in the private and the public sector. Because the GDPR applies in many different situations, many of its rules are rather vague and abstract. And the EU lawmaker did not specially consider the specific context of micro-targeting when drafting the GDPR. For example, freedom of expression and democracy play a larger role in the area of micro-targeting than in cases where, for instance, an app provider collects personal data for behavioural advertising.

More precise rules for personal data use for political micro-targeting may be needed (see also ICO, 2019). Perhaps the EU lawmaker could adopt rules for the use of personal data in the context of micro-targeting. However, adopting such rules would be difficult for the EU, as different EU member states have different traditions in the context of elections.

# FREEDOM OF EXPRESSION

Political micro-targeting is a form of political communication, and thus, is an exercise of the right to freedom of expression, which is guaranteed by both Article 11 of the EU Charter of Fundamental Rights, and Article 10 of the European Convention on Human Rights (ECHR). To understand the protection afforded to political micro-targeting as a form of political speech, we must turn to the case law of the European Court of Human Rights, the court that ultimately decides whether a restriction of freedom of expression is consistent with the ECHR.

## POLITICAL MICRO-TARGETING AS POLITICAL SPEECH

While the European Court of Human Rights has not to date considered a case involving political micro-targeting, it has held that a closely-related form of political communication, a political party's paid-for political advertising on television during an election, is a form of political speech enjoying the highest level of protection under Article 10. The publication of information "with a view to influencing voters is an exercise of freedom of political expression", and this is so, "[i]rrespective of the fact that it [is] presented as a paid advertisement" (*TV Vest v. Norway*, 2008). Paid-for political micro-targeting, as a form of political advertising, is therefore a form of political speech under Article 10. That conclusion is consistent with the Court's broad notion of what constitutes an exercise of freedom of expression, which includes: posting comments online during an election period (*Savva Terentyev v. Russia*, 2018), posting pictures on Instagram targeting public figures (*Einarsson v. Iceland*, 2017), uploading political videos to YouTube (*Mariya Alekhina v. Russia*, 2018), posting links to online videos targeting political parties (*Magyar Jeti Zrt v. Hungary*, 2018), distributing election leaflets *Andrushko v. Russia*, 2010), and displaying political posters (*Kandzhov v. Bulgaria*, 2008).

Indeed, the court has held that a political party's mobile app allowing voters to anonymously share pictures of ballots was a protected form of freedom of expression (*Magyar Kétfarkú Kutya Párt v. Hungary*, 2018). The court applied its well-established principle that Article 10 not only applies to the content of information expressed, but also to the *means of transmission*, and the *form* in which they are conveyed. The court has also held that people must be able to choose, without unreasonable interference from the government, the form they consider the most effective to reach a maximum number of people (*Women On Waves v. Portugal*, 2009). The political party's app had a communicative value, allowing voters to share information, and therefore constituted political expression under Article 10.

There is an important consequence of political micro-targeting being considered political speech, as such expression enjoys a 'privileged position' under Article 10 (*TV Vest v. Norway*, 2008). Because of that privileged position, the court applies its highest standard of scrutiny - strict scrutiny - to any restriction on political speech. Because there is 'little scope' for restrictions on political speech, any restriction must be 'narrowly interpreted', and its necessity 'convincingly established' by the government (*Vitrenko v. Ukraine,* 2008*)*. Further, Article 10's protection of expression on matters of public interest includes expression which is offensive, shocking or disturbing (*Dichand v. Austria*, 2002). It is also 'particularly important' that during the pre-election period opinions and information of all kinds are permitted to circulate freely (*Bowman v. UK*, 1988). Given the protection afforded to political speech, it is not surprising that when the court considered Norway's ban on paid political advertising on television, as applied to a Norwegian political party in the run-up to local elections, the court unanimously found a violation of Article 10 (*TV Vest v. Norway*, 2008). In sum, micro-targeting is a form of political expression that receives considerable legal protection.

## POLITICAL PARTIES AND ONLINE PLATFORMS

When considering restrictions on political micro-targeting, the Article 10 rights of a number of different actors are at issue, including an election candidate's freedom of expression (*Otegi Mondragon v. Spain,* 2011), a political party's freedom of expression (*Magyar Kétfarkú Kutya Párt v. Hungary*, 2018), an online platform's freedom of expression (*Cengiz v. Turkey*, 2015), and, indeed, the public's (voters') right to receive information (*Magyar Helsinki Bizottság v. Hungary*, 2016).

First, where a politician engages in political micro-targeting, this is an exercise of the politician's Article 10 right to freedom of expression and to impart information to potential voters. For decades, the court has recognised that while freedom of expression is important for everybody, it is 'especially so' for politicians, as they represent the electorate, and defend the electorate's interests. As such, interferences with a politician's freedom of expression are subject to the 'closest scrutiny' by the court (*Castells v. Spain*, 1992). Accordingly, the margin of appreciation (or the space and deference the court grants national authorities and courts) for assessing the 'necessity' of the penalty imposed on a politician is 'particularly narrow' (*Otegi Mondragon v. Spain*, 2011) (see Brems, 2019). Further, Article 10 protects a politician's expression in the context of a political debate, even where it only has a 'slim factual basis', and politicians are fully entitled to engage in exaggeration, and strong, polemical language (*Arbeiter v. Austria*, 2007).

Second, a political party's freedom of expression extends beyond the content of its political expression, but also extends to the *means* of transmission, including the mere making available of a mobile app to allow voters to anonymously share their voting ballots. The case establishing this principle was *Magyar Kétfarkú Kutya Párt v. Hungary*, where the court considered a fine imposed by Hungary's National Election Commission on a small political party for operating a mobile app enabling voters to anonymously share comments and photographs taken of their ballot papers during a 2016 referendum. Before the court, the Hungarian government argued that there had been no interference with the political party's freedom of expression, as the party had only provided a mobile app for voters, and had not engaged in political expression itself.

However, the court unanimously rejected the government's argument, and held that making the app available was an exercise of the political party's freedom of expression, and fully protected under Article 10. The court found that there had been a violation of the party's freedom of expression, as the government had failed to demonstrate how the secrecy or fairness of the referendum had been impacted by the app.

The court has also linked the importance of protecting a political party's freedom of expression to democracy itself. Because political parties' activities form part of a *collective* exercise of freedom of expression, this in itself entitles political parties to seek the protection of Article 10 (*United Communist Party of Turkey v. Turkey*, 1998). Further, political parties represent different shades of opinion to be found within a country's population, and by relaying this range of opinion, political parties make an immense contribution to political debate, which is at the very core of a democratic society. The court highlights the 'primordial role' played by political parties, emphasising that they are the 'only bodies which can come to power and have the capacity to influence the whole national regime' (*Oran v. Turkey*, 2014).The court also emphasises the unique value of political parties, in that they put forward proposals for an overall societal model before the electorate, and by their capacity to implement those proposals once they come to power, political parties differ from other organisations which intervene in the political arena.

Third, online platforms also enjoy freedom of expression. The European Court has indeed highlighted the importance of online platforms (such as Facebook, YouTube, and Instagram) for freedom of expression. For example, according to the court, YouTube is a 'unique' and an 'undoubtedly' important platform for political speech and political activities, with the court recognising that "political content ignored by the traditional media is often shared via YouTube" (*Cengiz and Others v. Turkey*, 2015). Similarly, in relation to Instagram, the court has emphasised that the internet plays an important role in enhancing the public's access to news and facilitating the dissemination of information in general (*Einarsson v. Iceland*, 2017). And platforms which facilitate the creation and sharing of webpages within a group enjoy the protection of Article 10, as they constitute a means of exercising freedom of expression (*Ahmet Yıldırım v. Turkey*, 2012).

In relation to Google's and other online platforms, the court has held that Article 10 guarantees freedom of expression to 'everyone', and it makes *no distinction* according to the nature of the aim pursued, or the role played by natural or legal persons in the exercise of that freedom. The internet is one of the principal means by which individuals exercise their right to freedom of expression, and provides "essential tools for participation in activities and discussions concerning political issues" (*Ahmet Yıldırım v. Turkey*, 2012). Further, the court has held that the operators of the file-sharing platform The Pirate Bay (allowing users to share copyright-protected digital material), were entitled to Article 10 protection, as they put in place the "means for others to impart and receive information within the meaning of Article 10", as Article 10 guarantees freedom of expression to everyone, and "[n]o distinction is made in it according to whether the aim pursued is profit-making or not" (*Neij v. Sweden*, 2013).

While the court has not to date considered the regulation of online political advertising, it has delivered a number of judgments on the regulation of political advertising in broadcasting. The most relevant judgment is *TV Vest v. Norway*, where a Norwegian political party argued that a ban on political advertising on television, during the run-up to elections, violated its right to freedom of expression. The court found a violation of Article 10. The court recognised that there could be relevant reasons for a ban on political advertising, such as preventing the 'financially powerful' from obtaining an 'undesirable advantage' in public debates, and 'ensuring a level playing field in elections'. The court was thus signalling that there are circumstances where it may accept regulation of political advertising is permissible on certain policy grounds.

However, the court held that the political party at issue, a small pensioners' party, was 'hardly mentioned' in election television coverage, and paid advertising on television became 'the only way' for it to put its message to the public. Moreover, the party did not fall within the category of a party that the ban was designed to target, namely financially strong parties which might gain an 'unfair advantage'. Thus, the court held that the general 'objectives' of the ban could not justify its application to the political party, and thereby violated its right to freedom of expression under Article 10. Thus, the Article 10 principles protecting political expression, and a political party's expression, in addition to the Court's judgment in *TV Vest*, would seem to suggest that a ban on online political micro-targeting would be difficult to reconcile with Article 10.

However, there is some uncertainty in the case law, as the court held in *Animal Defenders International v. UK (*2013) that a ban on paid political advertising on television in the UK did not violate Article 10. But unlike *TV Vest*, the case concerned an animal rights group (not a political party), which sought to broadcast a political advertisement *outside* an election period. For the first time under Article 10, the court held that a certain type of regulation, which the court called 'general measures', can be imposed 'consistently with the Convention', even where they 'result in individual hard cases' affecting freedom of expression. The court laid down a three-step test for determining whether a 'general measure' is consistent with Article 10: the court must assess (a) the 'legislative choices' underlying the general measure, (b) the 'quality' of the parliamentary review of the necessity of the measure, and (c) any 'risk of abuse' if a general measure is relaxed.

The court then applied its general-measures test to the ban on political advertising on television in the UK: first, the court examined the 'legislative choices' underlying the ban, and accepted that it was necessary to prevent the 'risk of distortion' of public debate by wealthy groups having unequal access to political advertising; and due to 'the immediate and powerful effect of the broadcast media'. Second, with regard to the quality of parliamentary review, the court attached 'considerable weight' to the 'extensive pre-legislative consultation', referencing a number of parliamentary bodies which had examined the ban. Third, as regards the risks from relaxing a general measure, the court held that it was 'reasonable' for the government to fear that a relaxed ban (such as financial caps on political advertising expenditure) was not feasible, given the 'risk of abuse' in the form of wealthy bodies 'with agendas' being 'fronted' by social advocacy groups, leading to uncertainty and litigation. Therefore, the court held that the total ban on political TV advertising was consistent with Article 10.

It is not clear whether the European Court of Human Rights would apply *Animal Defenders* to a law prohibiting online political micro-targeting. The judgment resulted in a divided Court (9-8 vote), and the court did not expressly overrule *TV Vest*. But it does signal that the court will accept, in some circumstances, that outright bans on political advertising may be consistent with freedom of expression, in order to prevent the risk of distortion of public debate by wealthy groups.

## NATIONAL RULES ON POLITICAL ADVERTISING

What rules are currently in force in Europe concerning online political advertising? We briefly outline the rules in France, Germany, the Netherlands and the UK, which represent widely divergent approaches.

At one end of the spectrum, is France, where Article L. 52-1 of the Electoral Code prohibits, during the six months prior to an election, "the use, for the purpose of election propaganda, of any commercial advertising in the press or any means of audiovisual communication". This rule also covers online public communication (Granchet, 2017).

Further, in late 2018, France introduced new rules under Art. L. 163-1 providing that in the three months prior to elections, online platforms must provide users with information about who paid for the "promotion of content related to a debate of general interest". Moreover, users must be provided with fair, clear and transparent information on the use of personal data in the context of the promotion of information content related to a debate of general interest. These rules have led some platforms, such as Twitter, to ban all political campaigning ads and issue advocacy ads in France (Twitter, 2019a). Similarly, Microsoft bans all ads in France "containing content related to debate of general interest linked to an electoral campaign" (Microsoft, 2019). Google also banned all ads containing "informational content relating to a debate of general interest" between April and May 2019 across its platform in France, including YouTube (Google, 2019). The French law led Twitter to even block an attempt by the French government information service attempting to pay for sponsored tweets for a voter registration campaign in the lead-up to European parliamentary elections (BBC, 2019). And in late 2019, Twitter introduced a global ban on paid-for promotion of political content on its platform (Twitter, 2019b), and Google implemented a new global rule limiting election ad audience targeting to age, gender, and general location (Google, 2019b).

Of course, platforms' bans may not capture all types of *indirect* political advertising that might take place, where ad campaigns do not promote a certain party or candidate, but the subject matter and message would favour certain candidates and parties because of their aligned agenda. At least the French law tries to capture all paid content "related to a debate of general interest", and not just campaigning and issue advocacy ads.

In Germany, under Article 7(9)(1) of the *Rundfunkstaatsvertrag* (RStV), paid political advertising is prohibited in broadcasting in an effort to prevent individual social groupings and forces from exerting a disproportionate influence on public opinion by purchasing advertising time (Etteldorf, 2017). Importantly, during elections, certain broadcasters are obliged to allocate free airtime to political parties for election advertising. The regulation of political advertising online depends not only on the online service itself but also on its provider. German law distinguishes between broadcasting and 'telemedia'. The transmission of a linear programme according to a schedule (especially live streaming services) via the internet is classified as broadcasting, and is therefore subject to the political advertising ban. Telemedia content (roughly speaking: internet content), on the other hand, is governed by Articles 54 et seq. of the RStV. Election advertising via on-demand audiovisual media services is prohibited under Article 58(3)(1), in conjunction with Article 7(9) of the RStV and, in other telemedia, must be separated from other content, in accordance with Article 58(1) of the RStV (Etteldorf, 2017). However, these rules do not apply to social media platforms like Facebook and YouTube.

At the other end of the spectrum is the Netherlands, where there are no specific restrictions concerning the type of political content that can be broadcast during elections; and Dutch law does not specifically regulate online political advertising during elections and referenda. In practice, political parties have limited budgets in the Netherlands. The Dutch government has proposed a new Political Parties Act, including new transparency obligations for political parties with regard to digital political campaigns and political micro-targeting (see Van Hoboken et al., 2019).

Finally, in the United Kingdom, paid political advertising in broadcasting is prohibited under the Communications Act of 2003. However, the ban does not apply online. While paid political advertising is not specifically restricted online through regulation, the UK Electoral Commission has emphasised that election spending rules "cover the costs of placing adverts on digital platforms"; and include the "costs of distributing and targeting digital campaign materials or developing and using databases for digital campaigning". Further, the Commission has recommended a number of reforms to election laws applicable to online political advertising, including (a) election and referendum adverts on social media platforms should be labelled to make the source clear; and (b) campaigners should be required to provide more detailed and meaningful invoices from their digital suppliers to improve transparency (Electoral Commission, 2018).

At the EU level, the European Commission has recognised some of the concerns related to online political micro-targeting, including that it creates increased possibilities to target citizens often in a 'non-transparent' manner, and may involve the processing of personal data of citizens 'unlawfully in the electoral context' (European Commission, 2018a). The Commission has introduced a self-regulatory code, and also guidance for member states about elections to the European Parliament. The self-regulatory Code of Practice on Disinformation, agreed with platforms including Facebook, Google and Twitter, includes that the platform will ensure transparency about political and issue-based advertising, also with a view to enabling users to understand why they have been targeted by a given advertisement (European Commission, 2018b).

Further, as the Commission notes, Article 14 of the Treaty on European Union provides that the European Parliament is to be composed of representatives of the Union's citizens; however, the procedure for the elections to the European Parliament is governed by national provisions in each member state, and national authorities are in charge of monitoring the elections at the national level (European Commission, 2018a). The Commission recommended that member states should encourage the disclosure of information on campaign expenditure for paid online political advertisements, including "information on any targeting criteria used in the dissemination of such advertisements" (European Commission, 2018a).

However, a controversy erupted in the run-up to European Parliament elections in 2019. Facebook implemented new rules on political advertising on its platforms, where any political party, candidate, group or individual within the EU were required to go through an ad authorisation process when planning to run ads related to politics or issues of national importance. The rules included that advertisers could only run ads in the country in which they are authorised (Facebook, 2019). This led political parties contesting the European parliamentary elections to criticise Facebook over the rules, arguing that Facebook's rules "prevent pan-EU parties from posting online political adverts across the 28-member bloc" (Khan, 2019).

Indeed, a letter was sent from three EU institutions to Facebook, which urged Facebook to amend its rules to allow European institutions, political groups, Members of the European Parliament, European political parties, to run pan-European advertising activities (Welle, Tranholm-Mikkelsen, & Selmayr, 2019). Notably, three weeks later, Facebook announced that it had "implemented temporary exemptions for the main Facebook pages of the European Parliament, European political groups and European political parties" (Kayali, 2019). However, there is no publicly available list of exempted groups, parties and EU institutions. In conclusion, national countries differ widely in how they regulate political advertising.

For completeness sake, we make a few brief remarks about self- and co-regulation. Platforms such as Facebook are implementing transparency mechanisms, including publicly-searchable political ad libraries. But while ad libraries make it much more difficult to post 'dark ads' (messages only visible to the targeted group [Hall Jamieson, 2018]), the ad libraries' "present implementations leave much to be desired" (Leerssen et al., 2019). For instance, definitions of 'political' ads "vary greatly and continue to raise significant line-drawing and enforcement challenges". Second, it is an open and difficult question whether and how platforms should ensure that ad buyers are properly identified. Third, so far platforms do not give much information on how and to whom political ads are targeted (Leerssen et al., 2019).

More generally, it is debatable whether national governments in the EU should leave the protection of democratic debate to online platforms. Should governments step in with regulation, akin to the regulation of political advertising in broadcasting? Should national governments rely upon online platforms themselves to ensure political micro-targeting is not damaging democracy? Do online platforms have the expertise to weigh the different interests involved, including the interest in free and fair elections?

## CONCLUSION

We discussed how online political micro-targeting is regulated in Europe. Political micro-targeting has a unique risk of harm not associated with traditional political advertising: interference with the rights to privacy and data protection. We showed that the GDPR generally applies to online political micro-targeting. The GDPR offers useful rules to protect privacy-related and other interests. For instance, the GDPR makes it more difficult for parties to buy detailed data about consumers. In the micro-targeting context, the GDPR is a necessary but not a sufficient protection.

We showed that political parties and online platforms have a strong freedom of expression claim in the context of political advertising. Political speech, including political advertising, deserves considerable protection in the case law of the European Court of Human Rights. Nevertheless, under certain conditions, national lawmakers can limit political speech. Therefore, national lawmakers could probably ban online political micro-targeting, at least for a period leading up to elections.

A complicating factor is the tension between EU-level regulatory action, and national rules in the area of election regulation. The EU has never stepped into the regulatory domain of national election regulation, as evidenced by the omission of rules on political advertising under the EU's rules for broadcasting (Audiovisual Media Services Directive), and general advertising (Unfair Commercial Practices Directive). Election regulation involves a particularly complex balancing of interests, and is tied to national culture and political history. Moreover, the EU has competence to regulate personal data, and to regulate elections for the European Parliament, but no specific competence to regulate national elections. Hence, national parliaments seem best placed to regulate political micro-targeting. Meanwhile, platforms themselves are deciding that the commercial reputational damage associated with allowing political micro-targeting may outweigh the commercial benefits (e.g., Microsoft, Twitter, and Google).

In this paper we discussed mostly what lawmakers in the EU (and four member states) do and *can* do in the area of micro-targeting. There are a range of possibilities, ranging from not regulating micro-targeting at all, to banning micro-targeting during certain periods. In between those two extremes there are many options, including rules that aim for more transparency. More debate and research are needed on what lawmakers *should* do.